# Ukrainian Arts and Humanities research in Scopus: A Bibliometric Analysis

**Serhii Nazarovets[1*], Olesya Mryglod[2]**

[1] Borys Grinchenko Kyiv University, 18/2 Bulvarno-Kudriavska Str., 04053 Kyiv, Ukraine
https://orcid.org/0000-0002-5067-4498

[2] Institute for Condensed Matter Physics of the National Academy of Sciences of Ukraine,
1 Svientsitskii St., 79011 Lviv, Ukraine
https://orcid.org/0000-0003-4415-7061

**Abstract**

**Purpose** – This article presents the results of a quantitative analysis of Ukrainian Arts and Humanities (A&H) research from 2012 to 2021, as observed in Scopus. The overall publication activity and the relative share of A&H publications in relation to Ukraine's total research output, comparing them with other countries. The study analyzes the diversity and total number of sources, as well as the geographic distribution of authors and citing authors, to provide insights into the internationalization level of Ukrainian A&H research. Additionally, the topical spectrum and language usage are considered to complete the overall picture.

**Design/methodology/approach** – This study uses the Scopus database as the primary data source for analyzing the general bibliometric characteristics of Ukrainian A&H research. All document types, except Erratum, were considered. A language filter was applied to compare the bibliometric characteristics of English versus non-English publications. In addition to directly imported data from Scopus, the study employs the ready-to-use SciVal tools to operate with A&H subcategories and calculate additional bibliometric characteristics, such as Citations per Publication (CPP), Field-Weighted Citation Impact (FWCI), and journal quartiles. Information on the country of journal publishers and details on delisted journals from Scopus were obtained from the official Source Title List available on the Elsevier website and the SCImago Journal & Country Rank Portal.

**Findings** – According to our results, the publication patterns for Ukrainian A&H research exhibit dynamics comparable to those of other countries, with a gradual increase in the total number of papers and sources. However, the citedness is lower than expected, and the share of publications in top-quartile sources is lower for 2020-2021 period compared to the previous years. The impact of internationally collaborative papers, especially those in English, is higher. Nevertheless, over half of all works remain uncited, probably due to the limited readership of the journals selected for publication.

**Originality/value** – This study provides original insights into the bibliometric characteristics of Ukrainian A&H publications between 2012 and 2021, as assessed using the Scopus database. Our findings reveal that Ukraine's A&H publications have higher visibility than some Asian countries with similar population sizes. However, in comparison to other countries of similar size, Ukraine's research output is smaller. We also discovered that cultural and historical similarities with neighboring countries play a more significant role in publication activity than population size. Our study highlights the low integration of Ukrainian A&H research into the global academic community, evident through a decline in papers published in influential journals and poor citedness. These findings underscore the importance for authors to prioritize disseminating research in influential journals, rather than solely focusing on indexing in particular databases.

## 1. Introduction

As early as 1948, Ukrainian-American linguist George Yurii Shevelov, writing under the pseudonym S. Yurskyi, expressed the following view: "The world turns to Ukrainian science when it is interested in internal Ukrainian issues; and that he is far more often interested in questions of a wider order – pan-Slavic or Eastern European, and these questions have not been worked out by Ukrainian science, then there is nothing to be surprised that in such cases the world turns to those scientific works that raise these questions" (Yurskyi, 1948). In our century Wiljan van den Akker (2016, p. 26) wrote something very similar: "Writing only in Dutch about Dutch poetry, will be absolutely the best guarantee that the world stays ignorant about the subject". These remarks highlight a dilemma that is widely discussed in the context of Arts and Humanities (A&H). On the one hand, it is natural to expect these disciplines to focus on locally nested problems (Petr *et al.*, 2021; Sivertsen, 2016). On the other hand, Pasteur wrote: "Scientists have a country; science has none." (Von Gizycki, 1973), emphasizing the universal nature of science, particularly in terms of knowledge sharing. Even a locally nested problem can be intriguing as a case study contributing to the global understanding. Furthermore, the analytical methods and tools employed can be adapted not only to investigate a similar local problem on the other side of the globe but also to other disciplines.

Thus, scientific communication serves as the foundation for the entire edifice of science. Within this context, the "local versus international" dilemma is closely linked to the issue of language selection. On one hand, scientists may prefer to use their native language for effective communication with a domestic audience, particularly for interdisciplinary matters; for developing research terminology in the national language, and for writing popular science works and textbooks primarily targeting the domestic market (Amano *et al.*, 2016). On the other hand, a lingua franca must be employed to make research findings accessible for foreign readers, share study results with the widest possible audience, and receive constructive feedback from peers. Moreover, journals indexed in core citation databases, which mainly include English-language publications (Albarillo, 2014; Liu, 2017), have better impact and visibility because scientists use these tools and metrics to find relevant peer-reviewed papers, whereas journals publishing articles in the vernacular appeal to a smaller readership (Dinkel *et al.*, 2004; Sanz-Casado *et al.*, 2021).

The presentation and dissemination of research findings embody a nuanced duality, particularly within the A&H disciplines. This is due to numerous researchers conducting locally-nested research and publishing their results in local journals, which significantly differ from international journals. Consequently, accurate evaluation of visibility and impact of these works on the scientific community becomes challenging (Sanz-Casado *et al.*, 2021). The goal of this study is to quantify the Ukrainian research output in A&H based on Scopus data and perform country-wise comparison. The focus on Ukraine presents a compelling and relevant case within the A&H context, given the country's rich linguistic and cultural heritage.

This is complemented by Ukraine's distinctive political landscape, as it expressed a strong preference for European integration after gaining its independence. Consequently, Ukraine seeks to assert its presence on the global stage, including the representation of its national research contributions across various A&H disciplines to the wider scientific community. However, achieving this goal within the A&H domain poses inherent challenges, as authors and readers actively utilize regional channels of scientific communication, unlike their counterparts in the natural sciences (Franssen and Wouters, 2019).

Examining how a multi-ethnic state with a diverse historical and cultural heritage addresses this issue is of great interest. Furthermore, the Russian-Ukrainian military confrontation, which began in 2014 after Russia annexed Crimea, has had a significant impact on Ukrainian humanities scholars. They have experienced specific social upheavals that undoubtedly influence the nature of humanities research conducted in the country and the broader region. Hence, this study aims to provide a unique bibliometric record of the transformative processes that have unfolded in recent years within the Ukrainian A&H field. Moreover, it will significantly contribute to a deeper understanding of contemporary humanitarian challenges in Eastern Europe.

The research questions addressed in this work are as follows:

*RQ1. What are the general bibliometric characteristics of Ukrainian output in the field of A&H as observed through commonly used international databases, which serve as data sources for quantitative description of research?*

The quantitative results are of particular interest to Ukrainian science policymakers and managers, as they oversee national assessment procedures and develop requirements for academic stuff. These results can also contribute to cross-country studies in A&H.

*RQ2. How are Ukrainian A&H research represented on an international scale, particularly among English-speaking colleagues?*

The answer to this question contributes to the existing body of research that explores various aspects of language usage in research, especially in the field of A&H.

The structure of this paper is as follows: a brief overview of research assessment peculiarities worldwide, with a specific focus on Ukraine, is presented in the section 'Background'. The section 'Data and Methods' outlines the sources and characteristics of the publication dataset used in our research. The main results are presented and discussed in several subsections of the 'Results and interpretations' section. Finally, the 'Discussion and Conclusions' section includes a summary and a discussion of the research limitations.

## 2. Background
*2.1 Bibliometrics in the Arts & Humanities*

The A&H research is heterogeneous, and scientists in these disciplines use different types of publications and languages (Blidstein and Zhitomirsky-Geffet, 2022; Kellsey and Knievel, 2004; Melchiorsen, 2019; Nederhof, 2006; Yang and Qi, 2021). The most popular citation databases, such as Web of Science (WoS) or Scopus, which are usually exploited as more reliable data sources for bibliometric analysis, are characterized by poorer coverage of Humanities. Therefore, they cannot be used to obtain representative data samples for comprehensive evaluations of this field (Archambault *et al.*, 2006; Borrego *et al.*, 2023; Kulczycki *et al.*, 2020; Mongeon and Paul-Hus, 2016). As a result, the bibliometric community rightly treats the use of quantitative assessment methods in the humanities with great caution (Hug and Ochsner, 2014; Ochsner *et al.*, 2017; Pedersen *et al.*, 2020).

However, despite these limitations, scientometrics has great potential for quantitatively describing scientific disciplines, including A&H, if the analysis is performed correctly and the interpretation takes into account the contexts. In particular, bibliometric analysis can be successfully used by scholars to reveal the typical publication patterns at the levels of journals, individual authors, or research institutions, to explore collaboration patterns (Donthu *et al.*, 2021; Kwiek, 2021; Vílchez-Román *et al.*, 2020; Wijewickrema, 2022), or to study the spread of innovative technologies (Agarwal *et al.*, 2022; Liu *et al.*, 2019; Su *et al.*, 2019). In recent years, scholars specializing in scientometrics from diverse countries have extensively conducted bibliometric analyses on national research output within the domain of A&H, utilizing essential core citation databases (for instance: Ardanuy *et al.*, 2022; Bui Hoai *et al.*, 2021; Golub *et al.*, 2020; Mohsen, 2021; Vlase and Lähdesmäki, 2023). Moreover, the results of such research performed in the context of a particular country (see, e.g., Mryglod *et al.*, 2021) can be used to improve national regulations of scientific activity or to perform cross-country analysis. Finally, there is a need to audit relevant information contained in popular databases, as they often serve as the first lens for creating the image of the national Art and Humanities field for the global scientific community (Márquez and Porras, 2020).

*2.2 Context of Ukraine*

Web of Science (WoS) and Scopus data are used in Ukraine to assess the scientific productivity of humanitarians on par with scientists from all other scientific disciplines. The Ukrainian national research evaluation is rather unsystematic and heavily relies on simple quantitative indicators, without considering the specific context of each discipline. This is particularly important for A&H (Hladchenko, 2022; Nazarovets, 2022). Despite the significance attributed to quantitative data and the unprecedented reliance of Ukrainian officials on indicators derived from these key citation databases, a comprehensive quantitative analysis of Ukrainian humanitarian studies has not yet been conducted.

As one of the countries with a small research output (see, e.g. Ayan *et al.,* 2023), Ukraine rarely appears on a map of cross-country scientometric research (e.g., Ma *et al.*, 2022; Wang, 2023). However, a few studies by Ukrainian scientometricians focusing on certain issues of the functioning and development of Ukrainian science, with particular attention to

Social Sciences and Humanities (SSH), can be found. For example, Kavunenko *et al*. (2006) conducted a comparative analysis of journals in the Social Sciences and Humanities of Ukraine and the world. Hladchenko and Moed (2021) examined the role of national journals in promoting research results at the local level, which also helps in meeting specific requirements of the Ukrainian national scientific policy. The large-scale bibliometric analysis of the Ukrainian Economics discipline which represents SSH, based on Scopus and Crossref data was performed by Mryglod *et al.*, 2021 and Mryglod *et al.*, 2022.

### 3. Data and methods

We chose to use the Scopus database as a data source for this study because it is one of the two most popular citation databases utilized for bibliometric analysis. Additionally, Scopus provides more consistent chronological coverage of Ukrainian humanities publications compared to WoS databases. Furthermore, while relevant publications can be found in both the specialized database Arts & Humanities Citation Index (AHCI) and the relatively new database Emerging Sources Citation Index (ESCI) on the WoS Core Collection platform, the Ukrainian national subscription only offers back files starting from 2015 for ESCI. Consequently, conducting an analysis based on WoS Core Collection data would lead to a rapid increase in the number of publications starting in 2015, which is an artifact of content indexing in the database rather than a true reflection of the publishing behavior of Ukrainian authors.

The following search query was used in Scopus to collect the data: 'AFFILCOUNTRY (Ukraine) AND SUBJAREA (ARTS) AND PUBYEAR < 2022 AND PUBYEAR > 2011'. All types of documents except Erratum were taken into account. Thus, we attribute to Ukraine all the documents, where at least one author's affiliation is related to Ukraine.

To compare the bibliometric characteristics of English vs non-English publications, a language filter was applied. In addition to analyzing data directly imported from Scopus, SciVal tools were used to operate with the subcategories of A&H and calculate additional bibliometric characteristics such as citations per publication (CPP), Field-Weighted Citation Impact (FWCI), and quartiles of journals. The official Source title list available on the Elsevier website[1] and the SCImago Journal & Country Rank Portal[2] were used to obtain information about the country of the journal publisher (since there is no information about the publishing country directly in Scopus) and the details about delisting the journals from Scopus. The starting year of journal indexing in Scopus and its Source-Normalized Impact per Paper (SNIP) score can be found on the journal profile pages in Scopus. The dataset was collected on February 7, 2023.

### 4. Results and interpretations

*4.1 Number of publications*
A total of **3,743** documents[3] published between 2012 and 2021, and indexed in the Scopus database for the "Arts and Humanities" subject area, are considered. It is interesting to note that this data sample has the potential to be larger, but approximately one-third of documents can be 'lost' for Ukraine due to the incompleteness of scholarly metadata (Mryglod and Nazarovets, 2023).

The Ukrainian output in A&H is only partially visible via Scopus, but a similar situation can be observed for other countries (Kulczycki *et al.*, 2018). Nonetheless, these data reflect the overall picture of the disciplinary area as perceived by the global community. While absolute estimations of the number of publications may lack informative value without additional context, it is reasonable to conduct the country-wise comparisons. In other words, it makes sense to use the same imperfect instrument to draw general conclusions about the state of the discipline within different national research systems.

Another problem arises when selecting countries for comparison with Ukrainian A&H. The development of disciplines within this field can be highly influenced by various factors such as local history, culture, language, and more. It is impossible to find two countries with identical heritage. In this study, we aim to compare countries based on population similarity (as the size of a country can impact on publication statistics (Kulczycki *et al.*, 2018)) and geopolitical position. However, it is important to note that such a choice is always somewhat arbitrary. Therefore, we consider the closeness of countries to Ukraine in a list of countries sorted by population in 2020, as well as the diversity of their other characteristics, to determine the initial group of countries for further comparison:

- Ukraine (43.7 million) – developing[4] Eastern European Post-Soviet country;
- Canada (37.7 million) – developed bilingual non-European country;
- Poland (37.8 million) – developing Eastern European Country neighboring with Ukraine;
- Argentina (45.2 million) – developing non-European country;
- Spain (46.8 million) – developed European country;
- Iraq (40.2 million) – developing Asian country;
- Uzbekistan (33.5 million) – developing Asian Post-Soviet country.

Three countries – the principally dissimilar ones – are considered in some cases simply to add more global context:
- USA (331 million) – developed non-European English-speaking country, in TOP3 countries with the largest population;
- United Kingdom (67.9 million) – developed European English-speaking country;
- China (1439.3 million) – a developing Asian country, the World leader by population.

The second group joins several countries which have common borders (and, thus, partially share the historical context) with Ukraine:
- Hungary (9.7 million);
- Romania (19.2 million);
- Slovakia (5.5 million);
- Poland (37.8 million).

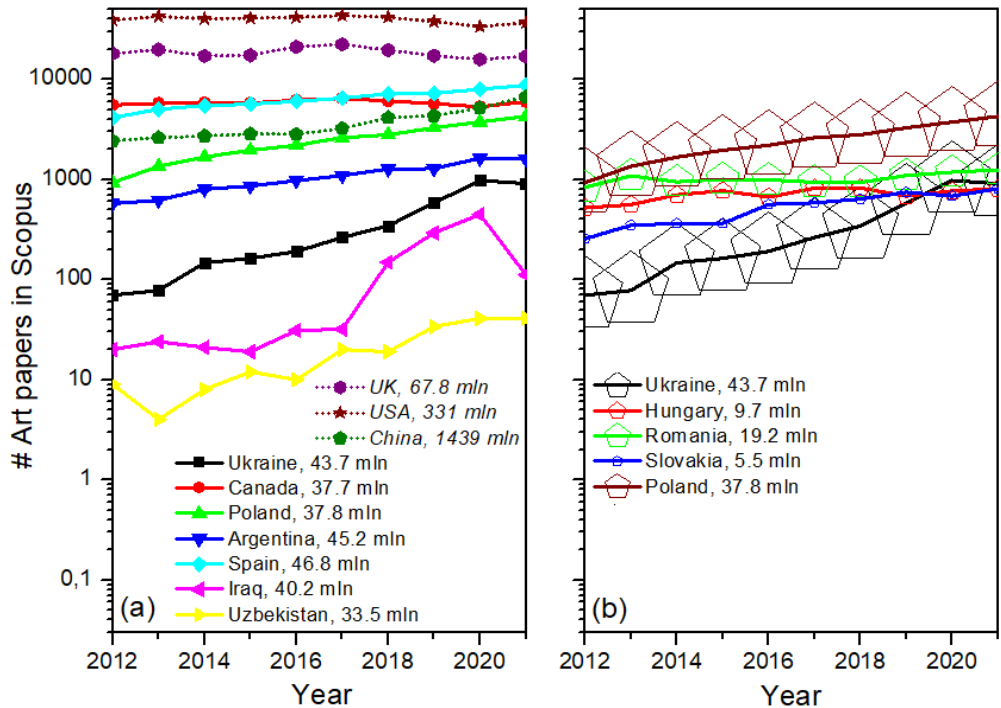

*Figure 1. The annual numbers of A&H publications in Scopus by authors from various countries: (a) countries closest to Ukraine by population but characterized by various economic and geopolitical positions plus English-speaking larger countries (USA and UK), and China, the largest country representing the Eastern part of the World; (b) countries that share a common border with Ukraine and, therefore, hypothetically have more cultural and historical similarities.*

The annual numbers of A&H outputs related to countries from the first and second groups are shown in Fig. 1. It can be observed that Ukraine is more visible via Scopus lenses compared to the Asian countries but less visible when compared to other countries of similar 'caliber', Fig. 1a. The Ukrainian output is also smaller than the output related to neighboring countries. In this context, one could speculate that the larger internal community of experts implies a lesser dependency on external audiences (Kulczycki *et al.*, 2018). However, the number of publications for Poland is higher not only compared to slightly larger Ukraine but also compared to much smaller countries (Fig. 1b).

While the absolute numbers differ among countries (sometimes varying by orders of magnitude), similar dynamics can be found. There is a gradual increase in the number of A&H publications by Ukrainian authors. This increase, characterized by varying rates, can also be seen for countries such as Poland, Spain, and Argentina. On the other hand, the outputs related to the United Kingdom, USA, and Canada, which share the common English language at least partially, as well as Romania and Hungary, exhibit rather constant annual values. It is evident that the increasing patterns cannot be solely explained by the global growth of scientific literature (Fortunato *et al.*, 2018; Price, 1963). To determine whether the share of A&H in the total output is increasing, the relative numbers are calculated next.

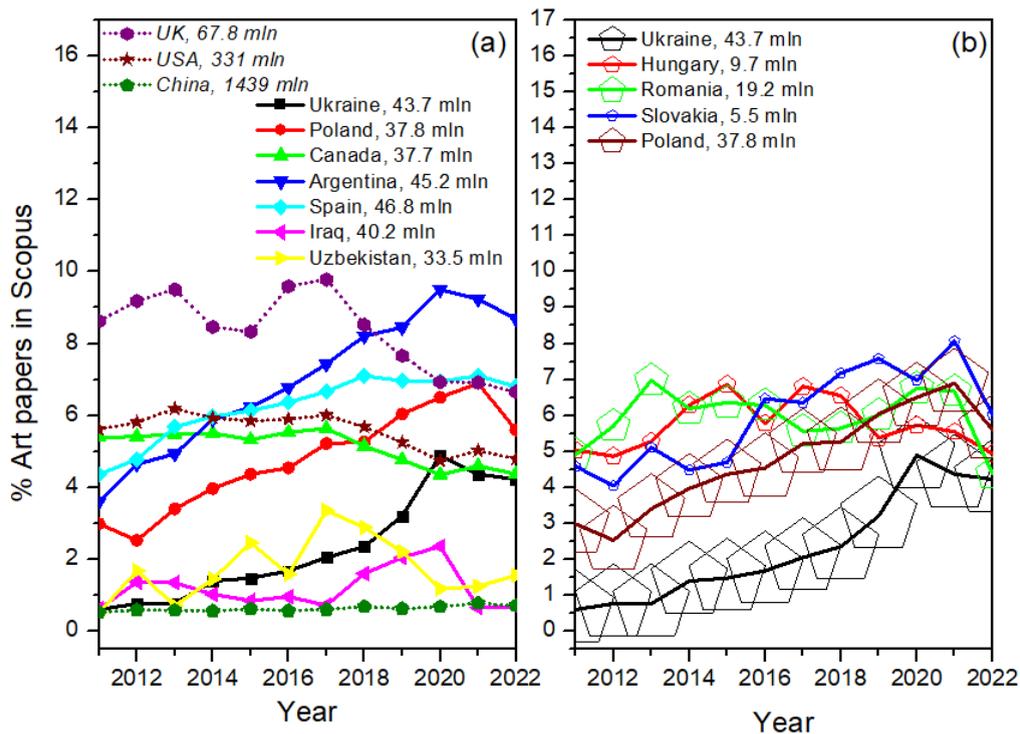

*Figure 2. The annual shares of A&H publications in Scopus by authors from different countries: the list of the countries for (a) and (b) panels is the same as in Figure 1.*

According to Scopus data, A&H papers accounted for approximately 2.7% of Ukraine's total scholarly output from 2012 to 2021. Country-wise comparisons of the annual values for the same groups of countries can be made, see Fig. 2. It can be observed that not only the absolute values but also the shares of A&H publications increased in Ukraine, rising from 1% in 2012-2014 to around 4.5% in 2020-2021. A similar dynamic, almost doubling, was observed for Poland and Argentina, while other countries showed a slower increase or even a decrease in relative numbers.

It is difficult to distill the separate factors that impact the observed tendencies. Clearly, the motivation of Ukrainian authors in A&H to publish their results in the sources indexed in international databases increases, even when compared to other disciplines. This may be a consequence of the gradual shift in the culture of scholarly communication as well as the changing rules of the national research evaluation system (e.g., since 2014, there has been an encouragement to publish results in foreign sources, and since 2018, a minimum number of publications visible in Scopus and/or Web of Science is required).

The following question arises: What approaches do Ukrainian authors adapt to achieve desired visibility in international databases? One way is to make greater efforts to publish results in already indexed sources, predominantly foreign ones. Another way is to promote national journals to expand the available space for Ukrainian publications. To examine the situation, an analysis at the level of source titles is conducted.

*4.2 Sources*

According to Scopus data, A&H papers by Ukrainian authors published between 2012 and 2021 can be found in 761 Sources. As shown in Figure 3 (left vertical scale), the number of different sources containing Ukrainian papers has been constantly increasing. The most remarkable increase in such sources" for Ukrainian A&H was observed in 2017 (the number of sources increased by almost 50%) and 2019 (another 33% increase). The countries with the TOP20 sources[5] (see Table 1) containing the largest number of Ukrainian A&H papers for each year between 2012 and 2021 were examined. Figure 3 (right vertical scale) reveals that the number of Ukrainian sources used for publication by Ukrainian authors doubled in two subsequent years after 2018. Evidently, the growth of Ukrainian A&H output is realized through an increase in papers by Ukrainian authors in foreign editions, as well as the inclusion of new Ukrainian source titles in the Scopus database (such as journals published in Ukraine; book editions, where at least one Ukrainian editor is found; conference proceedings, where Ukrainian institution is one of the organizers).

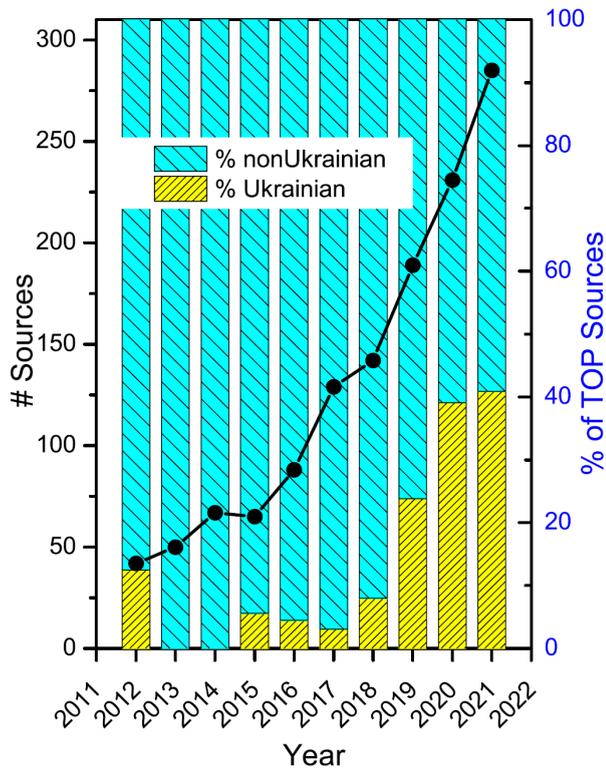

*Figure 3 Annual statistics for (i) the number of source titles containing A&H papers by Ukrainian authors (left vertical axis and line plot) and (ii) the share of source titles published in Ukraine in respect to the TOP20 most frequent ones (right vertical axis and bar diagram) between 2012 and 2012.*

While indexing in authoritative international databases is sometimes considered a guarantee of the worldwide visibility of a scientific journal, it can be shown that this is not a sufficient indicator of internationality. The latter is rather defined by the language of publications and their topical spectrum. In turn, the characteristics of authors and readers indicate the wideness of the audience. This aspect is especially crucial for the A&H area, where the already mentioned "local versus international" dilemma exists (Sivertsen, 2016; Petr *et al.*, 2021). The deep analysis of the internationality of sources is beyond the scope of this paper, but some preliminary results are shown in Table 1. The variations of two indicators of national geographical orientation of a journal discussed in (Moed, 2020) are used here. The shares of papers published by authors belonging exclusively[6] to the most contributing country are calculated for each journal (column 6). The geography of the citing authors is analyzed in the same way: the share of citing documents, where all authors exclusively belong to a single country – a leader by the number of citing publications – is calculated for each journal (column 8).

One can see that over one-half of all papers in most cases (highlighted by the bold font in Table 1) are not internationally collaborative and originate from a single country. It is possible to find definitions of local or domestic journals that are based on an even lower threshold for the number of papers authored by local authors (e.g., the value of 33% is mentioned in (Petr *et al.*, 2021)). While some journals in Table 1 can be referred to as more domestic due to the dominant share of papers exclusively from the home country (e.g., "Ukrainian Geographical Journal" for Ukraine or "Analele Universitatii din Craiova – Seria Stiinte Filologice, Lingvistica" for Romania), others are targeted at countries other than their own (e.g., "Stratum Plus", "Rusin", "Bylye Gody" which are defined as Moldavian or Slovakian, but predominantly publish papers by Russian authors). Similarly, one can speculate about the nationally oriented readership of the journals: a significant portion of citing documents can be attributed to a single country.

It should be noted that not all journals are represented in Scopus with complete archives. Furthermore, one journal has been discontinued in Scopus due to "Publication Concerns".

**Table 1.** Description of the TOP20 A&H sources with the highest number of publications by Ukrainian authors in the Scopus database within the period 2012-2021[7]. *The corresponding country names are highlighted in bold if the values in columns 6 or 8 exceed 50%.*

| Scopus Sources | # of [2012-2021] papers by Ukrainian authors | SNIP (2021) [Scopus coverage years] | Country of Scopus source[8] | # of [2012-2021] papers in the journal's profile | # (%) of publications by authors exclusively from [the most contributing country] | # of citing publications | # (%) of citing publications by authors exclusively from [the most actively citing country] |
|---|---|---|---|---|---|---|---|
| 1 | 2 | 3 | 4 | 5 | 6 | 7 | 8 |
| Stratum Plus | 203 | 0.715 [2014-2022] | Moldova | 941 | 536 (57%) **[Russian Federation]** | 729 | 421 (57.8%) **[Russian Federation]** |
| Rusin | 155 | 0.758 [2011-2022] | Moldova | 623 | 323 (51.8%) **[Russian Federation]** | 415 | 230 (55.4%) **[Russian Federation]** |
| *10th International Conference on Advanced Computer Information Technologies, ACIT 2020* | 139 | - | - | - | - | - | - |
| Psycholinguistics | | 0.211 [2019-2022] | Ukraine | 165 | 131 (79.4%) **[Ukraine]** | 112 | 77 (68.8%) **[Ukraine]** |
| Sententiae | 134 | 0.533 [2015-2022] | Ukraine | 173 | 127 (73.4%) **[Ukraine]** | 50 | 42 (84%) **[Ukraine]** |
| Journal of the National Academy of Legal Sciences of Ukraine | 122 | discontinued [2020-2021] | Ukraine | 129 | 117 (90.7%) **[Ukraine]** | 155 | 105 (69%) **[Ukraine]** |
| Ukrainian Geographical Journal | 104 | 0.237 [2018-2022] | Ukraine | 116 | 94 (81%) **[Ukraine]** | 85 | 68 (80%) **[Ukraine]** |
| Shidnij Svit | 79 | 0.000 [2019-2022] | Ukraine | 89 | 78 (87.6%) **[Ukraine]** | 7 | 7 (100%) **[Ukraine]** |
| Codrul Cosminului | 66 | 0.236 [2013-2022] | Romania | 196 | 64 (32.7%) [Ukraine] | 63 | 15 (23.8%) [Ukraine] |

| Danubius | 59 | 0.102 [2013-2020] | Romania | 234 | 126 (53.8%) **[Romania]** | 36 | 12 (33.3%) [Romania] |
|---|---|---|---|---|---|---|---|
| Analele Universitatii din Craiova - Seria Stiinte Filologice, Lingvistica | 56 | 0.187 [2012-2021] | Romania | 361 | 210 (58.2%) **[Romania]** | 104 | 29 (27.9%) [Ukraine] |
| Bylye Gody | 55 | 1.161 [2012-2022] | Slovakia | 1356 | 1001 (73.8%) **[Russian Federation]** | 1173 | 712 (60.7%) **[Russian Federation]** |
| East European Journal of Psycholinguistics | 43 | 0.151 [2019-2022] | Ukraine | 78 | 38 (48.7%) [Ukraine] | 54 | 29 (53.7%) **[Ukraine]** |
| Astra Salvensis | 42 | 0.417 [2013-2022] | Romania | 1003 | 301 (30%) [Russian Federation] | 1109 | 546 (49.2%) [Russian Federation] |
| Manuscript and Book Heritage of Ukraine | 40 | 0.224 [2020-2022] | Ukraine | 41 | 40 (97.6%) **[Ukraine]** | 3 | 2 (66.7%) **[Ukraine]** |
| Logos (Lithuania) | 37 | 0.630 [2008-2022] | Lithuania | 833 | 722 (86.7%) **[Lithuania]** | 211 | 164 (77.7%) **[Lithuania]** |
| History of Science and Technology | 36 | 0.016 [2020-2022] | Ukraine | 49 | 34 (69.4%) **[Ukraine]** | 30 | 20 (66.7%) **[Ukraine]** |
| Kyiv-Mohyla Humanities Journal | 30 | 0.129 [2019-2022] | Ukraine | 41 | 29 (70.7%) **[Ukraine]** | 17 | 5 (29.4%) [Ukraine] |
| Cogito | 29 | 0.316 [2018-2022] | Romania | 196 | 71 (36.2%) [Romania] | 82 | 15 (18.3%) [Nigeria] |
| Studia z Filologii Polskiej i Slowianskiej | 27 | 0.644 [2011-2021] | Poland | 206 | 107 (51.9%) **[Poland]** | 48 | 25 (52.1%) **[Poland]** |

Similarly, as it is performed for particular journals, the national orientation of the entire publication set can be estimated. Out of 3,743 Ukrainian A&H publications, 3,051 (81.5%) are not internationally collaborative and, thus, can be called national in this respect. The proportion is inverted for the citing documents: only 1,313 (26.3%) out of 4,984 publications that benefited from Ukrainian A&H works are related exclusively to Ukrainian authors. However, further analysis of citation impact reveals that a significant share of citations is attracted by publications involving foreign co-authors.

Seven source titles (and the next edition of the proceedings of the conference ACIT, which can be considered as the 8th), listed in Table 1, can also be found in the list of the TOP20 sources by the number of citing documents:
- Quaternary International (United Kingdom)
- Stratum Plus (Moldova)
- Rusin (Moldova)
- CEUR Workshop Proceedings (USA)
- Journal of Physical Education and Sport (Romania)
- Quaternary Science Reviews (United Kingdom)

- 2021 11th International Conference on Advanced Computer Information Technologies, ACIT 2021
- Sententiae (Ukraine)
- Bylye Gody (Slovakia)
- Sustainability (Switzerland)
- Journal of Archaeological Science: Reports (Netherlands)
- Astra Salvensis (Romania)
- Sprawozdania Archeologiczne (Poland)
- Journal of the National Academy of Legal Sciences of Ukraine (Ukraine)
- Naukovyi Visnyk Natsionalnoho Hirnychoho Universytetu (Ukraine)
- Ido Movement for Culture (Poland)
- Ukrainian Geographical Journal (Ukraine)
- Estudios de Economia Aplicada (Spain)
- Journal of Advanced Research in Law and Economics (Romania)
- E3S Web of Conferences (France)

To assess not only the "direction" of impact but also its magnitude, the SNIP indicators can be used. This kind of indicator is used to measure a journal's contextual citation impact, taking into account the characteristics of the subject field (Moed, 2010). The distribution of Ukrainian A&H publications according to SNIP journal quartiles is shown in Figure 4. According to these results, although the absolute number of A&H publications by Ukrainian authors gradually increased during 2012-2021 in Scopus, the share of work published in the TOP journals (belonging to the 1st or 2nd quartiles) decreased in the last two years. To investigate the citation impact in more detail, a more granular subject classification is used in the next subsection.

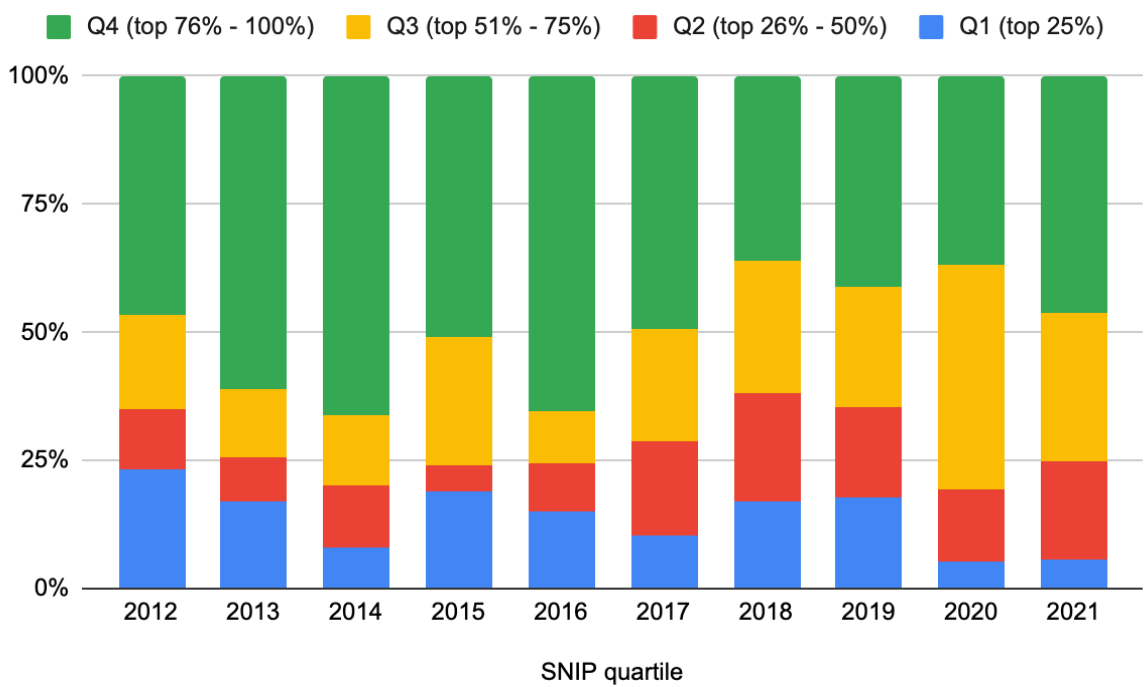

*Figure 4 The shares of Ukrainian A&H papers published in journals, differentiated by their SNIP quartiles in 2012-2021.*

*4.3 Citation Impact in respect of the Disciplinary Profile*

42% (1,573) of the 3,743 Ukrainian A&H publications are cited at least once. The share of non-cited papers is slightly higher (63%) for the dataset containing publications by Ukrainian authors only, and the annual tendency to have a larger share of cited publications if foreign authors are involved is preserved.

SciVal allows one to investigate the subject area of publications in more detail (see Table 2). One can see that a large portion of Ukrainian A&H publications visible via the Scopus database is represented by History (almost 43% of publications), followed by Language and Linguistics (26.4%) and Philosophy (17.8%). While some correlation between the number of outputs and the number of authors is expected, the proportion of authors versus documents is slightly higher for archaeology compared to the three leading subcategories mentioned before. It is possible to speculate about the differences in the methods of research: while it is expected to find historians or philosophers who work alone, teamwork is required to perform, say, archaeological expeditions. The last two columns of Table 2 give an idea of the impact of Ukrainian outputs. Curiously, besides the "Arts and Humanities (miscellaneous)" subcategory, the larger average number of citations per publication is found for "Conservation". However, it is meaningless to directly compare the citedness for different disciplinary subcategories, and FWCI indicator should be used instead. This metric is defined as the ratio of the total citations received by any given paper to the total citations that would be expected based on the average of that particular topic in the same period (Zanotto and Carvalho, 2021). "Conservation" appears to be the most impactful according to FWCI values (last column): the corresponding publications were cited 1.29 times more than expected. The citedness of publications within the rest of the subcategories of Ukrainian A&H is poorer than it potentially could be. The weakest citation impact is for 'Language and Linguistics' (FWCI=0.36).

**Table 2** Characteristics of publication statistics [2012-2012] related to Ukraine in respect of Scopus A&H subcategories. Several ASJC (All Science Journal Classification) codes can be assigned to the same publication here.

| Subcategory | Publications | Citations | Authors | CPP | FWCI |
|---|---|---|---|---|---|
| History | 1608 | 2232 | 2355 | 1.4 | 0.75 |
| Language and Linguistics | 989 | 819 | 1179 | 0.8 | 0.36 |
| Philosophy | 667 | 880 | 1075 | 1.3 | 0.91 |
| Archaeology (Arts and Humanities) | 585 | 1352 | 1115 | 2.3 | 0.45 |
| Literature and Literary Theory | 465 | 292 | 505 | 0.6 | 0.54 |
| General Arts and Humanities | 434 | 739 | 722 | 1.7 | 0.96 |
| Religious Studies | 331 | 247 | 458 | 0.7 | 0.60 |
| History and Philosophy of Science | 227 | 417 | 471 | 1.8 | 0.48 |
| Arts and Humanities (miscellaneous) | 127 | 668 | 436 | 5.3 | 0.72 |
| Visual Arts and Performing Arts | 113 | 155 | 183 | 1.4 | 0.88 |
| Conservation | 70 | 263 | 541 | 3.8 | 1.29 |
| Museology | 61 | 45 | 118 | 0.7 | 0.62 |
| Classics | 50 | 34 | 65 | 0.7 | 0.59 |
| Music | 34 | 39 | 63 | 1.1 | 0.97 |
| *Total in the Arts and Humanities subject area* | 3,743 | 6,525 | 6,226 | 1.7 | 0.68 |

To determine the subject areas most affected by Ukrainian A&H, we conducted a similar analysis of disciplinary categories and subcategories of the citing works. According to our findings, approximately one-third (27.9% to be precise) of citing publications are associated with Social Sciences; nearly one quarter (24.5%) are related to A&H. Following these disciplines are Earth and Planetary Sciences (7.5%), Computer Science (5.7%), Environmental Science (4.8%), Business, Management and Accounting (4.7%), Economics, Econometrics and Finance (4.3%). In terms of subcategories, History accounts for one-third (32.3.%) of the citing documents, while Archaeology comprises almost another one-third (27.9%). These subcategories are followed by Language and Linguistics (21.7%), and Philosophy (12.8%).

*4.4 Document type*

The variety of forms that can be used to publish the results in A&H is one of the features that make this area a very special object for scientometric analysis. It is known that publications in the form of Books or Book Chapters are more typical for the area of A&H compared to other disciplines. Still, the share of journal papers indexed in the Scopus database is much higher. This is expected due to the peculiarities of database coverage, differences in the process of articles and books writing, etc. The dominant share of publications within our dataset – 2651 (72%) – corresponds to journal articles. Other types of publications are significantly less numerous: review – 516, book chapter – 275, conference paper – 190, note – 40, editorial – 16, book – 15, letter – 4, and short survey – 1.

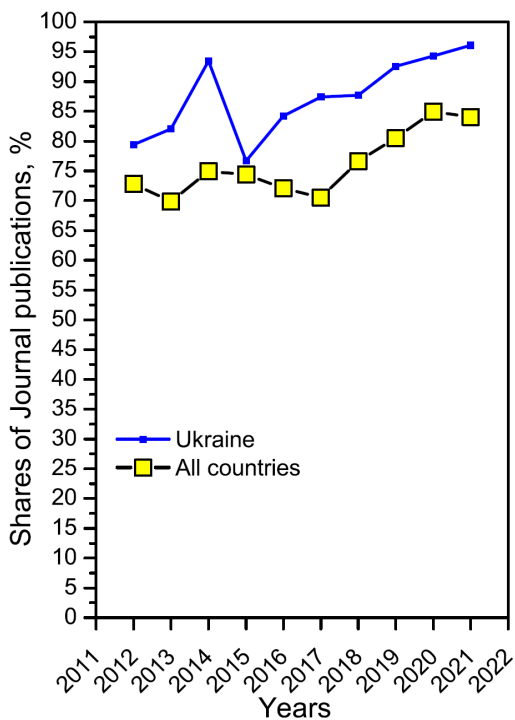

*Figure 5. The annual shares of A&H journal publications (Article, Review, Note, Editorial, Letter, Short Survey document types) published in 2012-2021 by authors from all countries (squares) and Ukraine (circles).*

Scopus' search for all papers within A&H subject category between 2012 and 2021 irrespectively of the authors' country allows one to draw a general conclusion: the proportion of works published in journals (Article, Review, Note, Editorial, Letter, Short Survey)[9] compared to those published in the form of books (Book Chapter, Book)[10] is about 2:1. This proportion cannot be considered characteristic of the A&H area in general due to the limitations of database coverage. For example, a smaller gap between the shares of journal and book publications is reported in (Kulczycki *et al.*, 2018), where the data from national scholarly databases for eight European countries are used. However, it is still reasonable to perform a country-wise comparison of publication patterns based on the same data source. In particular, Figure 5 shows that the annual share of journal publications in A&H, irrespective of the country of authors, gradually increases starting approximately from 2015-2017. The comparison of results for separate countries[11] revealed that this tendency holds for all countries, but for some, it is more evident, while for others, it is less noticeable (e.g., for Ukraine).

It is interesting to note that all A&H book publications by Ukrainian authors visible in Scopus are written not in Ukrainian, but rather in English, with a few in German.

*4.5 Authorship and Collaboration*

Art and Humanities papers published by 6,226 Ukrainian authors between 2012 and 2021 are visible via Scopus. The number of unique authors increases every year: over 10 times more authors were involved in 2021 compared to 2012. Typically, Ukrainian A&H publications are written by one author (51.9% of all publications). The average number of authors per paper is 2, but a few atypical documents are authored by the large groups (e.g., the reports for large projects: the most collaborative publication by 394 authors contains the report for one of the EC FP7 Projects[12]; the report for RESET project[13] authored by 192 of its participants). The averaged value can be considered less informative due to the disciplinary heterogeneity, as mentioned before. To give an example, 94.7% of works are published by 1–5 authors, and the majority of these publications (72.9%) are also related to Social Sciences. The other much smaller share (5.3%, only 199) of documents is characterized by larger co-authorship groups – these works are much less related to Social Sciences (44.7%). The annual dynamics of the average number of authors per publication can be observed in Fig. 6 (right vertical scale): if two highly-collaborative project reports mentioned above are considered, the corresponding average values for 2015 and 2016 are remarkably higher (circles), while the fluctuations in values are much smaller if just these two documents are removed from the analysis (dotted region and the corresponding average values for 2015 and 2016). One can see that a slow, but constant increase in average values is accompanied by an increase in multiple-author publications (see Fig. 6, left vertical scale and shaded areas).

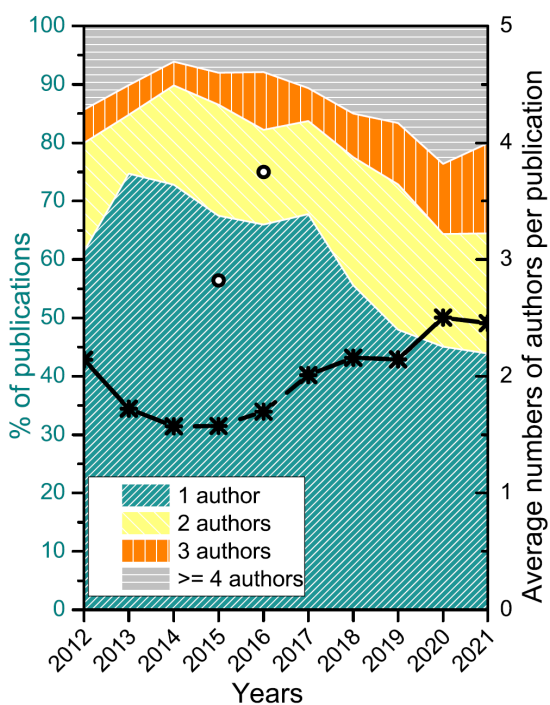

*Figure 6. The annual shares of Ukrainian A&H publications are categorized by the sizes of co-authorship teams, i.e. consisting of 1, 2, 3, or at least 4 authors (left vertical scale). The corresponding average numbers of authors per paper are indicated by black symbols (right vertical scale). Two publications by large co-authorship groups are considered outliers that disturb the average values (circles for 2015 and 2016) and are therefore omitted during averaging (black symbols and the dotted region).*

The impact of publications in respect to the size of the co-authorship team can be seen in Table 3. As mentioned before, the majority of A&H works are single-authored (and related only to Ukraine – this is how our data selection is performed), which is consistent with publishing traditions in the field. But at the same time, this category of publications has the lowest citation impact, which is more than twice lower compared to the expected value: FWCI = 0.38. According to the results, collaboration – whether institutional or at the national level – is associated with a higher citation impact. The most impactful works are published in co-authorship with foreign colleagues; in this case, the impact is higher than expected, FWCI = 1,52. Most of the joint papers of Ukrainian humanitarians were written in co-authorship with scientists from Russia (195), Poland (146), the United States (105), Germany (87), and the United Kingdom (65).

**Table 3.** The citedness of Ukrainian A&H publications [2012-2012] in relation to the characteristics of co-authorship, based on Scopus data.

|  | Publications | Percentage | Citations | CPP | FWCI |
|---|---|---|---|---|---|
| International collaboration | 624 | 16.7% | 3208 | 5.1 | 1.52 |
| Only national collaboration | 518 | 13.8% | 746 | 1.4 | 0.8 |
| Only institutional collaboration | 659 | 17.6% | 915 | 1.4 | 0.66 |
| Single authorship | 1942 | 51.9% | 1656 | 0.9 | 0.38 |

*4.6 Language*

According to Scopus data, two-thirds (63%) of Ukrainian A&H publications during 2012-2021 are in English (63%), while only 12.9% are in the official local language – Ukrainian. The TOP3 most-used languages apart from English are Russian (19.6%), Polish (1.5%), and German (1.1%). In the previous study, presented at the ISSI 2021 conference (Nazarovets and Mryglod, 2021), the usage of languages in publications by Ukrainian researchers in several subfields of Humanities indexed in WoS was investigated. To this end, the study compared Ukraine with other non-English-speaking countries in Eastern Europe that have similar post-Soviet backgrounds. Fig. 7 shows the country-wise comparison of language spectra. Similarly to Figs. 1-2, the comparison is performed within two groups: (a) countries with comparable population numbers (Argentina, Spain, Iraq, Uzbekistan, Poland) and (b) countries that share a common border with Ukraine (Poland, Hungary, Romania, Slovakia).

Although the dominant share of A&H publications in Scopus is in English, the important role of other languages, especially local national languages, can be seen in Fig. 7. It is interesting to note that the share of publications in English for Ukrainian data is more similar to neighboring countries. More variations are observed when considering countries of similar size. This may serve as another hint about the factors that impact publication patterns in A&H.

The results of a more granular analysis of Ukrainian publications at the level of subcategories of A&H are presented in Fig. 8. One can see that documents in English prevail in each subcategory of A&H, except for the subcategory Literature and Literary Theory, where the majority of publications are non-English.

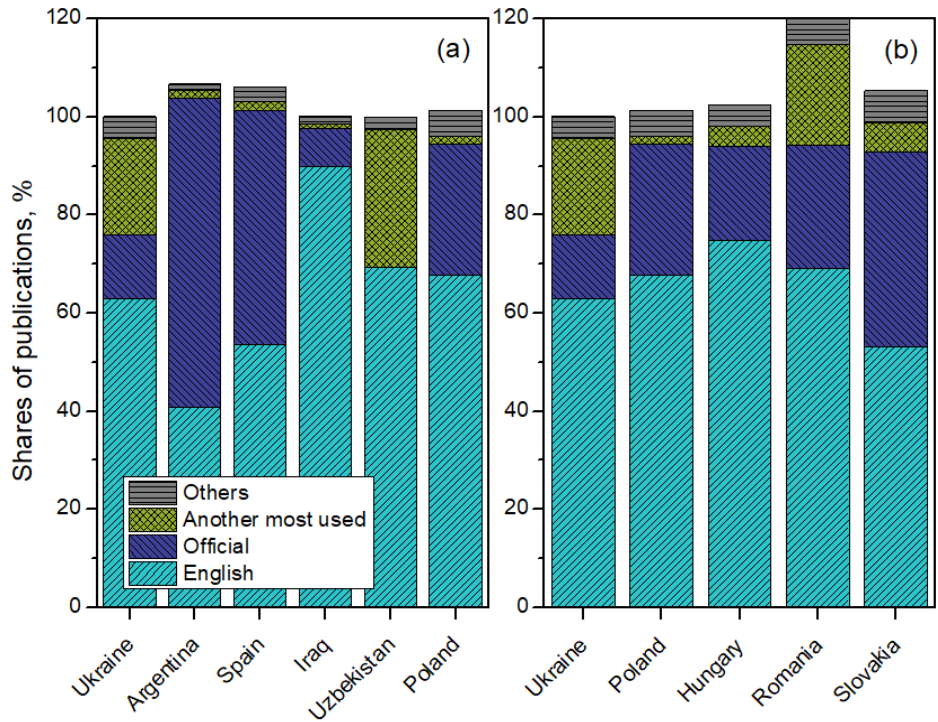

*Figure 7. The distribution of A&H publications in different languages (English, the official national language, another most used language, and the rest) for several countries: (a) Ukraine and countries with similar population sizes (Argentina, Spain, Iraq, Uzbekistan, Poland); (b) Ukraine and neighboring countries (Poland, Hungary, Romania, Slovakia) that share a common border. The total sum of shares may exceed 100% due to the language detection peculiarities in Scopus, where multiple languages can be attributed to a single document.*

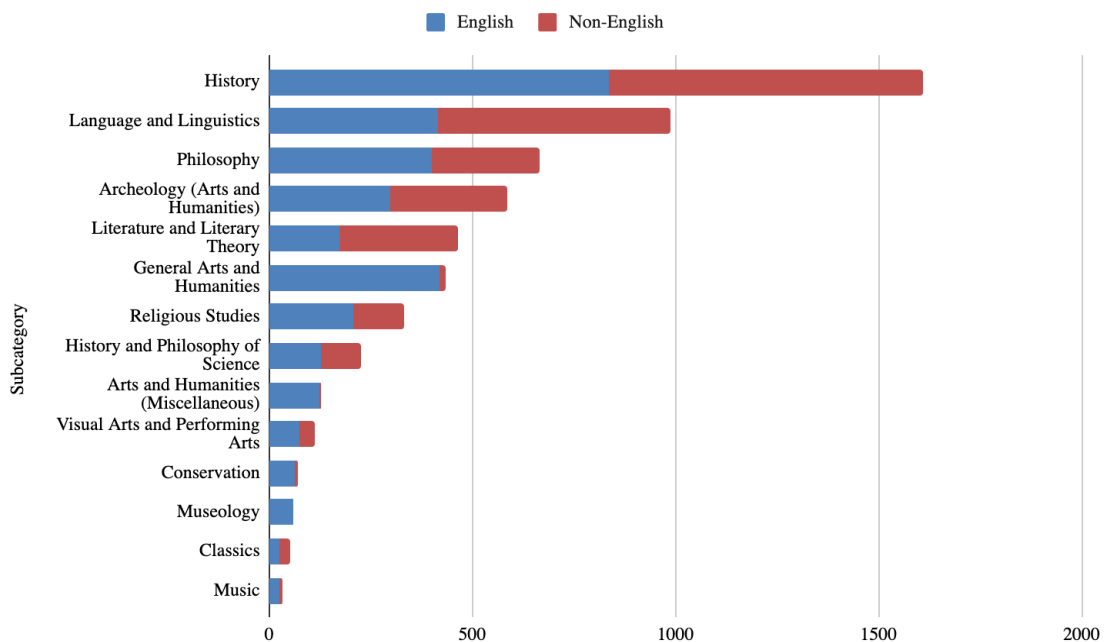

*Figure 8. The distribution of Ukrainian A&H publications across various disciplinary subcategories, categorized by language (English vs. non-English).*

To speculate on the correlations between language usage and the scholarly impact, FWCI values were calculated for English vs. non-English publications within A&H subcategories. According to these numbers, the citation impact of English publications is more than twice as large as publications in other languages for the majority of subcategories. The corresponding values of FWCI for the TOP5 most represented subcategories are as follows:
- History: i.e., FWCI = 1.07 for publications in English vs FWCI = 0.4 for publications in other languages;
- General Arts and Humanities: 0.97 vs 0.5;
- Language and Linguistics: 0.55 vs 0.23;
- Philosophy: 1.32 vs 0.3;
- Archaeology (Arts and Humanities): 0.65 vs 0.25.

In general, the citation impact for all non-English Ukrainian A&H publications, regardless of the disciplinary subcategory, is significantly below the global average.

## 5. Discussion

Our study reveals the main bibliometric characteristics of Ukrainian A&H publications between 2012 and 2021 as observed through the Scopus database. According to our findings, the visibility of Ukrainian A&H in this context is higher compared to some Asian countries with comparable populations. However, Ukrainian output is smaller when compared to other countries of a similar 'caliber' in terms of population. While the country's size is considered an important factor shaping publication activity in A&H, our results indicate that the absolute annual publication rates are closer for countries with more similar geopolitical positions, specifically those that share common borders with Ukraine, in our case. We speculate about the importance of cultural and historical similarities.

The observed increase in annual numbers of outputs is a characteristic of A&H not only for Ukraine: similar patterns are observed for Poland, Argentina, and China. Along with the global growth of scientific literature (Bornmann *et al.*, 2021), the expansion of the A&H segment is also found for Ukraine. This feature of Ukrainian scholarly output is not exotic, either. Of course, the explanations may differ for different countries. The increase in the absolute number of Ukrainian scholarly publications is accompanied by an increase in the relative share of A&H and the number of sources. In our belief, these tendencies are caused by a gradual evolution of publication practices and scholarly communication in general, as well as changes in the rules of rewarding in the research sphere. The expanding share of journal publications versus books can be considered as a side-effect of a desire to have more publications covered by international databases (Abramo *et al.*, 2023). The coverage of non-English language books in Scopus is poor (Giménez-Toledo *et al.*, 2017). Special attention to journal articles from national research evaluations can also lead to changes in publishing practices, as has already been found for other countries (Hammarfelt and Haddow, 2018; Kulczycki *et al.*, 2018).

The topical spectrum of Ukrainian A&H is studied: History is found to be the most represented subcategory of the field by the total number of publications (as well as the total number of unique authors and citations) in Scopus. The most-related disciplinary area is found to be Social Sciences: the majority of A&H publications are co-attributed to it, but this holds only for works published by 1–5 authors – the most typical size of co-authorship teams. Such a natural connection between A&H is confirmed also by the disciplinary distribution of citing works. However, when interpreting these results, it should also be remembered that different approaches to the disciplinary classification of papers can affect bibliometric analysis, which is confirmed by the results of other studies (Arhiliuc and Guns, 2023; Guns *et al.*, 2018).

In the course of data, several interesting nuances with publication metadata were noticed. For example, the source-wise method of subject categorization of records applied in Scopus resulted in the proceedings of the 10th International Conference on Advanced Computer Information Technologies (ACIT 2020) being included in the Arts and Humanities category. Hypothetically, only four contributions labeled as "Information Technologies in Historical Science" led to this categorization of the ACIT proceedings. As a consequence, 139 publications primarily related to Computer Sciences were automatically included in the Ukrainian A&H output for 2020. Despite the small sample size (which is not uncommon for A&H in Scopus or WoS), it was sufficient to make ACIT 2020 one of the most influential sources in 2020. Additionally, a distortion in the typical co-authorship pattern was observed, with an overrepresentation of publications co-authored by 5 or more authors. The expected number of publications for different sizes of co-authorship teams was determined using a random sampling technique. Other noteworthy outliers were the two reports from large collaborations mentioned earlier (EC FP7 Projects and RESET project), which were incorrectly labeled as regular articles.

One of the most important aspects of our results is the conclusion about the low integrity of Ukrainian A&H in the global scholarly information system. This conclusion can be drawn from different factors. Firstly, Ukrainian authors in A&H tend to publish their results in Ukrainian journals, which are often targeted at local audiences (10 out of the TOP20 most used sources are published in Ukraine, and for 7 of them, over half of both publications and citing documents are related exclusively to Ukrainian authors). Secondly, other sources most used by Ukrainian humanitarians are published in neighboring countries. Here, the strong influence of the Soviet past has to be taken into account: there is a long-standing tradition of publishing in Russian editions, and the Russian language is still widely used in Ukraine (although it is not the official language in Ukraine, it is the second most used language in Ukrainian A&H visible in Scopus; the largest number of collaborative works are published in co-authorship with Russian authors). Since 2014 publications in Russian journals are not encouraged due to the first stage of the Russian-Ukrainian war, but at least 4 out of the TOP20 most used sources, while not officially attributed to Russia, are largely targeted at Russian audiences according to both the shares of publications and citing publications related exclusively to Russian authors. Therefore, one can conclude that (i) the strong dependency on the Russian expert environment is still preserved in Ukrainian A&H and (ii) the internationalization of the field is formal in the sense that the results are published in sources with a narrow audience, and the majority of published results remain uncited, while the rest are usually cited less than expected, especially for non-English publications. The annual shares of papers in the highest-quartile journals have decreased, with the lowest numbers characterizing the output for 2020 and 2021.

Despite the contentious nature of utilizing citations as a means of evaluating research, recent scholarly investigations indicate a discernible association between research quality and citation metrics, even within the realm of humanities (Thelwall *et al.*, 2023). Consequently, the increasing number of Ukrainian A&H outputs in Scopus, coupled with the decreasing number of highly-influential source titles and poor citedness, may be considered as hidden signals about the wrong motivation of authors. Rather than prioritizing the best platform to disseminate their results, the main priority seems to have shifted towards the formal status of journals, such as indexing in Scopus, which does not necessarily imply influence or global visibility by default. This aligns with the conclusions of a previous study, which primarily focused on journals in the natural sciences: "…the practice used to reward scientists and institutions in Ukraine probably does not encourage Ukrainian scientists to seek the optimal channel for the presentation of their research outputs. Instead, most Ukrainian scientists are trying to quickly publish as many papers as possible" (Nazarovets, 2020). However, additional research, including authors interviews, is required to convincingly prove this claim.

To improve the visibility and impact of the work of Ukrainian scientists, it is recommended to conduct a complete audit and reform of the national system of scientific assessment in Ukraine, taking into account the best practices of other countries (Lewandowska *et al.*, 2022; Pedersen *et al.*, 2020; Pölönen *et al.*, 2021). In the process of evaluating authors and institutions working in A&H, Ukraine should abandon the simplistic use of quantitative indicators, or at least significantly reduce their weight in favor of independent expert assessment, following current trends in research evaluation (CoARA, 2022; Hatch and Curry, 2020; Hicks *et al.*, 2015; Wilsdon *et al.*, 2015). The updated Ukrainian system should consider the specifics of A&H, including the various channels for disseminating humanitarian knowledge and the use of different languages for publishing scientific results.

Additionally, we suspect that the full-scale Russian-Ukrainian war, which began in February 2022, and the reaction of the Ukrainian and global community to this invasion (Nazarovets and Teixeira da Silva, 2022; Van Noorden, 2023), will impact the structure of co-authorship and the language preferences of Ukrainian authors in the A&H fields.

## 6. Conclusions
The results of the analysis of Ukrainian A&H publications for 2012-2021 allow us to draw the following conclusions.

Firstly, we found that the absolute annual number of Ukrainian A&H papers has increased, accompanied by an increase in the number of published sources. While country size is an important factor in shaping publication activity in A&H, our results indicate that countries with geopolitical proximity to Ukraine tend to have similar annual publication rates, suggesting the influence of cultural and historical similarities.

Another interesting result is related to the topical spectrum of Ukrainian A&H: History appears to be the most represented subcategory in terms of total publications, unique authors, and citations.

An important aspect of our findings is the low integration of Ukrainian A&H into the global scholarly information system: Ukrainian A&H authors tend to publish their results in local journals targeting domestic audiences. Additionally, Ukrainian A&H continues to exhibit substantial reliance on the Russian expert environment.

One *limitation* of this study is that we only considered quantitative citation rates and did not analyze the sources of these citations, such as the presence of unethical manipulations that can significantly distort the overall results due to relatively small statistics. Citations to publications in the sources discontinued from Scopus due to violations of publishing ethics are also not distinguished here.

To obtain a more complete picture of the publishing activity of Ukrainian authors in the field of A&H, it is necessary to significantly expand the source base of the bibliometric analysis, particularly by adding more non-English-language local sources. In many countries, comprehensive bibliographic data can be obtained from national CRIS systems (Zhang and Sivertsen, 2020), but such a national system has not been developed for Ukraine thus far.

**Data availability**
The search results of the literature in Scopus underlying the bibliometric analysis are available as an open source set of Supplementary files at Zenodo
https://doi.org/10.5281/zenodo.7682823.

**Acknowledgements**
The authors would like to thank all Ukrainian defenders for the possibility to finalize and publish this work.

**Competing interests**
The authors do not have any competing interests.

**Authors' contributions**
The authors contributed equally to all aspects of the paper.

**Funding statement**

No funding was received by either author for this research.

**Notes**
1. https://www.elsevier.com/solutions/scopus/how-scopus-works/content. The latest update – January 2023.
2. https://www.scimagojr.com/
3. January 2023
4. According to https://www.worlddata.info/developing-countries.php Accessed in February 2023
5.  if the annual statistics is too poor to build a TOP20 rating, all sources with at least two papers are taken into account
6.  no international collaboration is allowed, unlike in the original paper (Moed, 2020)
7.  the data for this particular table were retrieved in January 2023
8. according to https://www.scimagojr.com/
9.  Scopus search query: SUBJAREA (arts) AND PUBYEAR < 2022 AND PUBYEAR > 2011 AND (LIMIT-TO (DOCTYPE, "ar") OR LIMIT-TO (DOCTYPE, "re") OR LIMIT-TO (DOCTYPE, "no") OR LIMIT-TO (DOCTYPE, "ed") OR LIMIT-TO (DOCTYPE, "le") OR LIMIT-TO (DOCTYPE, "sh"))
10. Scopus search query: SUBJAREA (arts) AND PUBYEAR < 2022 AND PUBYEAR > 2011 AND (LIMIT-TO (DOCTYPE, "ch") OR LIMIT-TO (DOCTYPE, "bk"))

11. the same list of countries as in Figures 1 and 2 was used for comparison
12. The Ocean of Tomorrow" Project CoCoNet funded by the EU's research funding program between 2007 and 2013, see
https://ec.europa.eu/search/index.do?QueryText=7th+Framework+Programme+%28FP7%29&op=Search&swlang=en
13. The RESET project (RESponse of humans to abrupt Environmental Transitions), a program of research funded by the Natural Environment Research Council (UK) between 2008 and 2013